\documentclass[copyright, creativecommons]{eptcs}
\providecommand{\event}{WWV 2012}

\usepackage{breakurl} 
\usepackage{graphicx,latexsym,amssymb,amsmath,alltt}
\usepackage{color}
\usepackage{algorithm}
\usepackage{algorithmic}

\definecolor{blank}{rgb}{0.7,0.7,0.7}

\newtheorem{theorem}{Theorem}[section]
\newtheorem{definition}[theorem]{Definition}
\newtheorem{example}[theorem]{Example}

\long\def\comment#1{}

\renewcommand{\phi}{\varphi}

\def\defemb#1#2{\expandafter\def\csname #1\endcsname
                              {\relax\ifmmode #2\else\hbox{$#2$}\fi}}
\defemb{cA}{{\cal A}}
\defemb{cB}{{\cal B}}
\defemb{cC}{{\cal C}}
\defemb{cD}{{\cal D}}
\defemb{cE}{{\cal E}}
\defemb{cF}{{\cal F}}
\defemb{cG}{{\cal G}}
\defemb{cH}{{\cal H}}
\defemb{cI}{{\cal I}}
\defemb{cJ}{{\cal J}}
\defemb{cL}{{\cal L}}
\defemb{cM}{{\cal M}}
\defemb{cN}{{\cal N}}
\defemb{cO}{{\cal O}}
\defemb{cP}{{\cal P}}
\defemb{cR}{{\cal R}}
\defemb{cS}{{\cal S}}
\defemb{cT}{{\cal T}}
\defemb{cU}{{\cal U}}
\defemb{cV}{{\cal V}}
\defemb{cW}{{\cal W}}
\defemb{cX}{{\cal X}}
\defemb{cZ}{{\cal Z}}

\makeatletter
\newenvironment{prog}{\vspace{1.0ex}\par
\obeylines\@vobeyspaces\tt}{\vspace{1.0ex}\noindent
}
\makeatother
\newcommand{\startprog}{\begin{prog}}
\newcommand{\stopprog}{\end{prog}\noindent}
\newcommand{\pr}[1]{\mbox{\tt #1}}   

\catcode`\_=\active
\let_=\sb
\catcode`\_=12
\makeatother

\title
{
 Using the DOM Tree for Content Extraction\thanks
 {
    This work has been partially supported by the Spanish \emph{Ministerio de Ciencia e Innovaci\'on} 
   under grant TIN2008-06622-C03-02 and by the \emph{Generalitat Valenciana} under grant PROMETEO/2011/052. 
   David insa was partially supported by the \emph{Ministerio de Educaci\'on} under grant FPU AP2010-4415.
 }
}
\author
{
  Sergio L\'opez \quad Josep Silva \quad David Insa
  \institute
  {
   Departamento de Sistemas Inform\'aticos y Computaci\'on\\
   Universitat Polit\`ecnica de Val\`encia
   E-46022 Valencia, Spain.}
  \email{\quad slopez@upv.es \quad jsilva@dsic.upv.es \quad dinsa@dsic.upv.es}
  }
\def\titlerunning{Using the DOM Tree for Content Extraction}
\def\authorrunning{Sergio L\'opez, Josep Silva \& David Insa}

\begin{document}
\maketitle

\begin{abstract}
The main information of a webpage is usually mixed between menus, advertisements, panels, and other not necessarily related information; and it is often difficult to automatically isolate this information. This is precisely the objective of \emph{content extraction}, a research area of widely interest due to its many applications. Content extraction is useful not only for the final human user, but it is also frequently used as a preprocessing stage of different systems that need to extract the main content in a web document to avoid the treatment and processing of other useless information. Other interesting application where content extraction is particularly used is displaying webpages in small screens such as mobile phones or PDAs. 
In this work we present a new technique for content extraction that uses the DOM tree of the webpage to analyze the hierarchical relations of the elements in the webpage. Thanks to this information, the technique achieves a considerable recall and precision. Using the DOM structure for content extraction gives us the benefits of other approaches based on the syntax of the webpage (such as characters, words and tags), but it also gives us a very precise information regarding the related components in a block, thus, producing very cohesive blocks. 
\end{abstract}

\section{Introduction}
\begin{figure*}[t]
	\centering
		\includegraphics[width=0.80\textwidth]{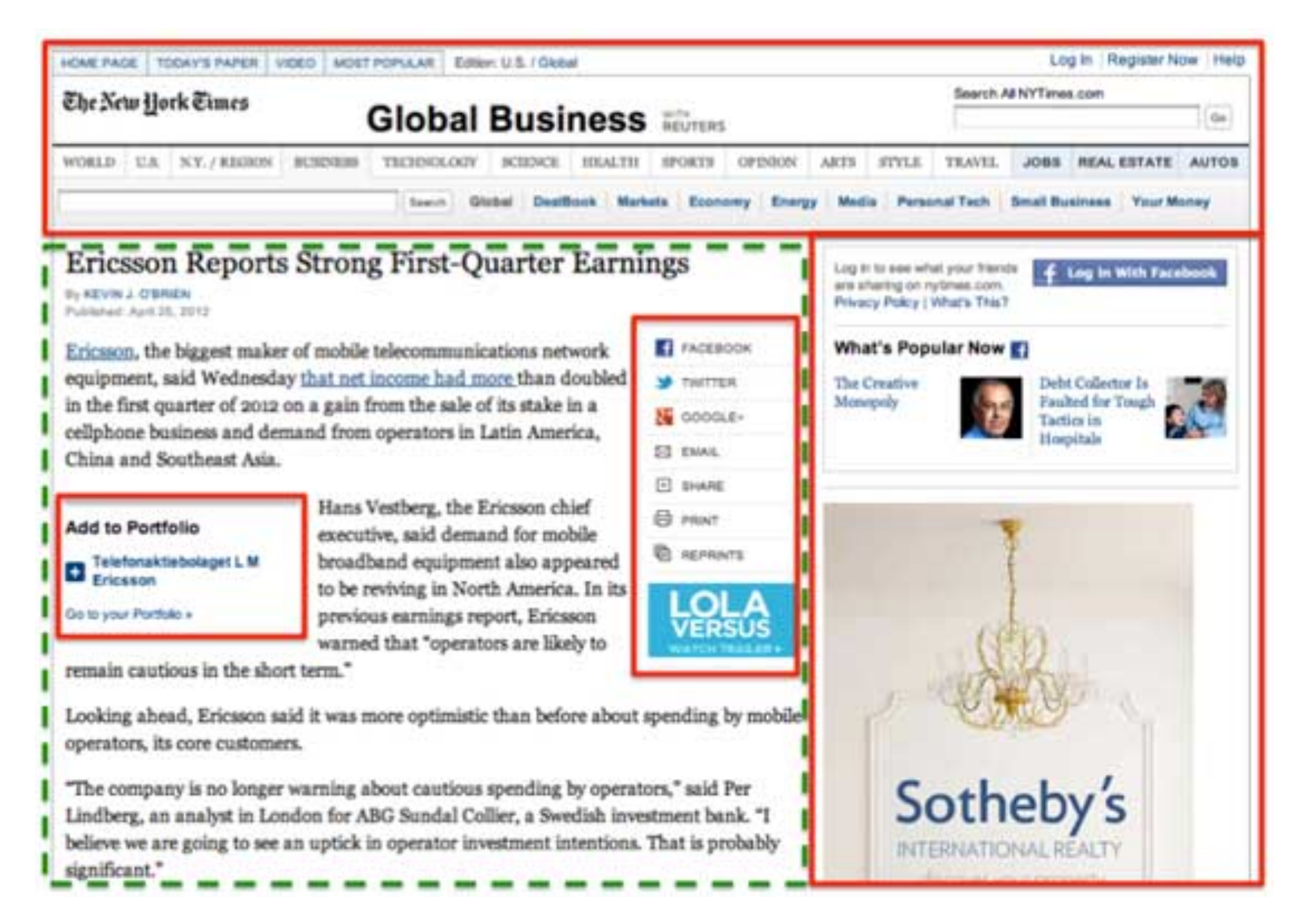}
	\caption{Blocks of a webpage from the New York Times website}
	\label{fig-blocks}
\end{figure*}

\emph{Content Extraction} is one of the major areas of interest in the Web for both the scientific and industrial communities. This interest is due to the useful applications of this discipline. Essentially, content extraction is the process of determining what parts of a webpage contain the main textual content, thus ignoring additional context such as menus, status bars, advertisements, sponsored information, etc. Content extraction is a particular case of a more general discipline called \emph{Block Detection} that tries to isolate every information block in a webpage. For instance, observe the blocks that form the webpage in Figure~\ref{fig-blocks}, and in particular, the main block delimited with a dashed line. Note that inside the main block there are other blocks that should be discarded.  

It has been measured that almost 40-50\% of the components of a webpage can be considered irrelevant \cite{Gib05}. Therefore, determining the main block of a webpage is very useful for indexers and text analyzers to increase their performance by only processing relevant information.   
Other interesting applications are the extraction of the main content of a webpage to be suitably displayed in a small device such as a PDA or a mobile phone; and the extraction of the relevant content to make the webpage more accessible for visually impaired or blind. 

Our technique combines ideas from other works such as \cite{Got08,Wen10}, and it also uses additional information that is 
explicit in the DOM tree of webpages, and that allows the technique to produce very accurate results.

In summary, the main advantages of our technique are the following:
\begin{itemize}
\item It does make no assumptions about the particular structure of webpages.
\item It only needs to process a single webpage (no templates, neither other webpages of the same website are needed).
\item No preprocessing stages are needed. The technique can work online. 
\item It is fully language independent (it can work with pages written in English, German, etc.).
\item The particular text formatting of the webpage does not influence the performance of the technique.
\end{itemize}

The rest of the paper has been structured as follows:
In Section~\ref{sec_rel} we discuss the state of the art and show some problems of current techniques
that can be solved with our approach. In Section~\ref{sec_DOM} we recall the DOM model and provide some useful notation. Then, we present our algorithms and explain the technique with examples in Section~\ref{sec_CE}. In Section~\ref{sec_impl} we give some details about the implementation and show the results obtained with a collection of benchmarks. Section~\ref{sec_CS} presents three interesting cases of study using our tool. Finally, Section~\ref{sec_concl} concludes.

\section{Related Work}\label{sec_rel}

Many different techniques have been proposed to solve the problem of content extraction. 
Some of them are based on the assumption that the webpage has a particular structure 
(e.g., based on table markup-tags) \cite{Li03}, that the main content text is continuous \cite{Ari09}, that the system knows a priori the format of the webpage \cite{Li03}, or even that the whole website to which the webpage belongs is based on the use of some template that is repeated \cite{KHG05}. This allows the system to analyze several webpages and try to deduce the template of the webpage in order to discard menus and other repeated blocks. 

The main problem of these approaches is a big loss of generality. In general, they require to previously know or parse the webpages, or they require the webpage to have a particular structure. This is very inconvenient because modern webpages are mainly based on \emph{$<$div$>$} tags that do not require to be hierarchically organized (as in the table-based design). Moreover, nowadays, many webpages are automatically and dynamically generated and thus it is often impossible to analyze the webpages a priori. 

There are, however, other approaches that are able to work online (i.e., with any webpage) and in real-time (i.e., without the need to preprocess the webpages or know their structure). One of these approaches is the technique presented in \cite{Got08}. This technique uses a \emph{content code vector} (CCV) that represents all characters in a document determining whether they are content or code. With the CCV, they compute a \emph{content code ratio} to identify the amount of code and content that surrounds the elements of the CCV. Finally, with this information, they can determine what parts of the document contain the main content.
Another powerful approach also based on the labeling of the characters of a document has been presented in 
\cite{Wen10}. This work is based on the use of \emph{tag ratios} (TR). Given a webpage, the TR is computed for each line with the number of non-HTML-tag characters divided by the number of HTML-tags. 
The main problem of the approaches based on characters or lines such as these two, or words such as \cite{Fin01}, is the fact of completely ignoring the structure of the webpage. Using characters or words as independent information units and ignoring their interrelations produces an important loss of information that is present and explicit in the webpage, and that makes the system to fail in many situations.

\begin{example}
\label{Ex-TagRatios}
Consider the following portion of a source code extracted from an IEEE's webpage:

{\scriptsize
\startprog
<body>
   (...)
   <div id="maincontent">
      <a name="Abstract"><h2>Abstract</h2></a>
      <p>Most HTML documents (...)
      (...)
      (...) delivers the best results.</p>
   </div> 
   <div id="footer">
      <p class="bottomstuff">Copyright IEEE. All rights are (...) </p>
      <p class="links">
      <a href="IEEE\_Xplore.html" target="blank">Help</a> | 
      <a href="/xpl/techform.jsp">Contact Us</a> | (...)
   </div>
   (...)
</body>
\stopprog
}

\begin{figure}[h]
       \vspace{-1cm}
	\centering
		\includegraphics[width=0.80\textwidth]{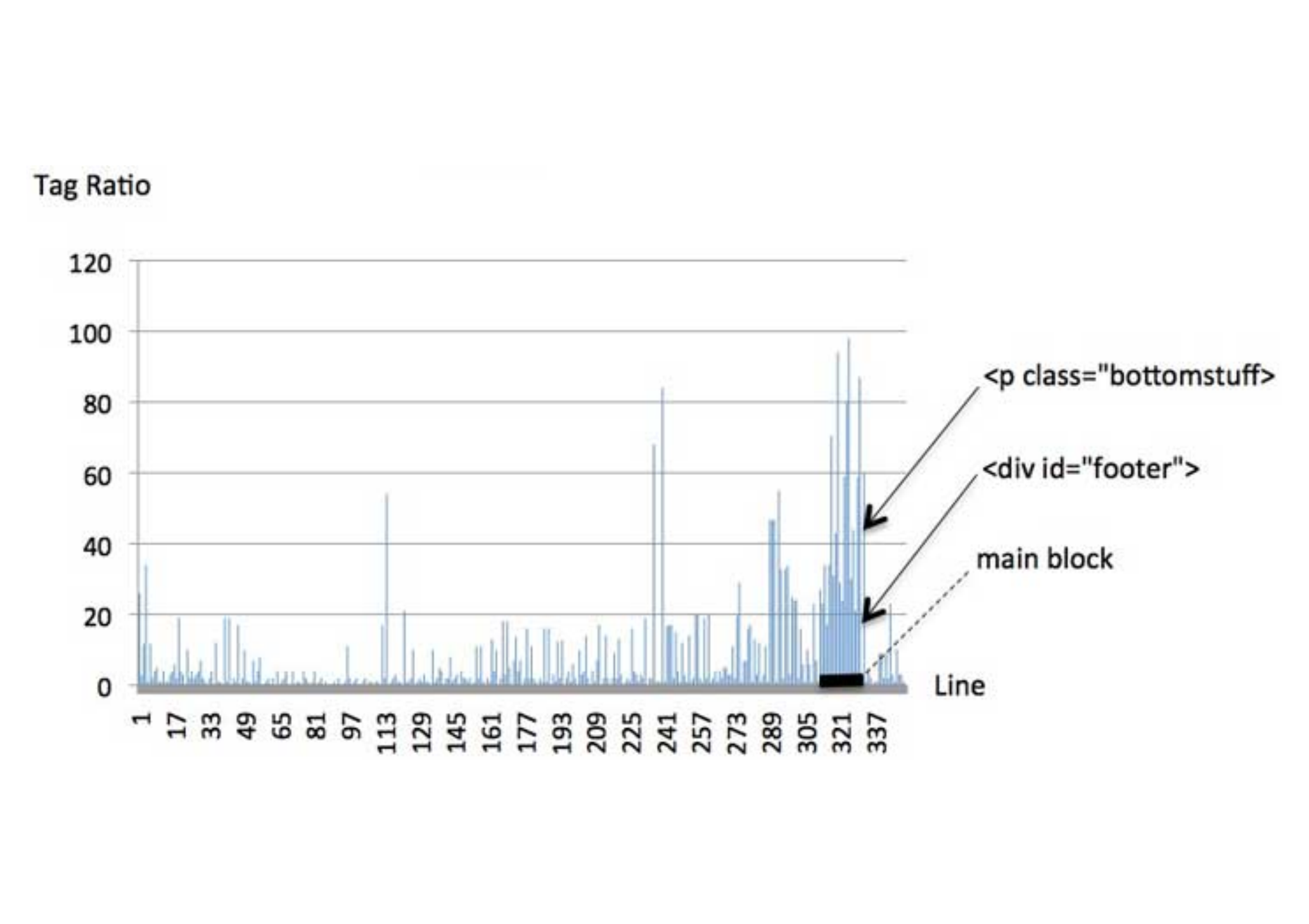}
       \vspace{-1cm}		
	\caption{Tag ratios}
	\label{fig:CETR}
\end{figure}

\noindent 
The tag ratios associated to this webpage are shown in Figure~\ref{fig:CETR}. Observe that
the initial part of the footer (which is not part of the main content) is classified as
part of the main content because it starts with a high tag ratio. Unfortunately, this method does not take into account the information provided by tags, and thus, it fails to infer that the footer text belongs to a different $<div>$ than the other text classified as relevant.

\end{example}

The distribution of the code between the lines of a webpage is not necessarily the one expected by the user. The format of the HTML code can be completely unbalanced (i.e., without tabulations, spaces or even carriage returns), specially when it is generated by a non-human directed system. As a common example, the reader can see the source code of the main Google's webpage. At the time of writing these lines, all the code of the webpage is distributed in only a few lines. In this kind of webpages tag ratios are useless.

In this work, we solve this situation by using a ratio similar to the tag ratio but based on the DOM structure of the webpage. This makes our approach keep the good properties of the tag ratios approach, but it also solves the problems shown in the previous example because the technique combines the computed ratios with the information of the DOM tree. In particular, because the DOM tree is independent of the distribution of the code between the lines of the HTML webpage, our technique is able to work with any webpage independently of how the webpage was generated or formatted.

It should be clear that---as it happens in the other appro\-aches---the technique could fail to detect the main block if other block (e.g., the footer) contains more text density that the real main block. But our technique easily distinguishes between different blocks (thanks to the DOM information), and 
it does not mix information from different blocks as in Example~\ref{Ex-TagRatios}.

Although essentially different to our work, there exist other techniques that make use of the DOM structure, and thus, they could exploit the same information than our technique. The most similar approach is the one presented in \cite{Gup03}. This approach presents a proxy server that implements a set of filters for HTML documents. These filters include HTML cleaning (e.g., removing images, scripts, etc.), HTML refactoring (e.g., removing empty tables), and deleting advertisements (e.g., with a blacklist of URLs that allow them to remove external publicity content).  Some of these transformations are used by our technique, but the objective is different, we do not want to clean, improve or transform the original webpage; our goal is to detect the main content and remove all the other components of the webpage. Also the implementation is different, our tool is not based on a proxy server; it is implemented in the client side, and thus it is independent of any external resource.  

There are some approaches specialized for a particular content such as tables that are somehow related to our work. They do not focus on block detection but in content extraction from tables \cite{Dal11}, or in wrappers induction \cite{Kus97,Coh02}. Other related approaches are based on statistical models \cite{Koh08,Koh09} and machine learning techniques \cite{Bal06,Gib07} and they use densitometric features such as link density and text features such as number of words starting with an uppercase letter \cite{Koh10}.

\section{The DOM tree} \label{sec_DOM}

The Document Object Model (DOM) \cite{DOM} is an API that provides programmers with a standard set of objects for the representation of HTML and XML documents. Our technique is based on the use of DOM as the model for representing webpages. Given a webpage, it is completely automatic to produce its associated DOM structure and vice-versa. In fact, current browsers automatically produce the DOM structure of all loaded webpages before they are processed.

The DOM structure of a given webpage is a tree where all the elements of the webpage are represented (included scripts and CSS styles) hierarchically. This means that a table that contains another table is represented with a node with a successor that represents the internal table. Essentially, nodes in the DOM tree can be of two types: tag nodes, and text nodes.\footnote{We make this assumption for simplicity of presentation. In the current DOM model, there are 12 types of nodes, including the type text.} Tag nodes represent the HTML tags of a HTML document and they contain all the information associated with the tags (e.g., its attributes). Text nodes are always leaves in the DOM tree because they cannot contain other nodes. This is an important property of DOM trees that we exploit in our algorithms.

\begin{definition}[DOM Tree] \label{Def-DOM}
Given an HTML document $D$, the DOM tree $t=(N,E)$ of $D$ is a pair with a finite set of nodes $N$ that contain either HTML tags (including their attributes) or text; 
and a finite set of edges $E$ such that $(n \rightarrow n') \in E$, with $n,n' \in N$ if and only if the tag or text associated with $n'$ is inside the tag associated with $n$ in $D$. The reflexive and transitive closure of $E$ is represented 
with $E^*$.
\end{definition}

For the purpose of this work, it does not matter how the DOM tree is built. In practice, the DOM's API 
provides mechanisms to add nodes and attributes, and provides methods to explore and extract information from the tree.

\begin{example}
\label{Ex-TagRatios}
Consider the following portion of a source code extracted from the entry ``information retrieval" at Wikipedia:
{\scriptsize
\startprog
<body>
   <h1 id="firstHeading" class="firstHeading"> Information retrieval</h1> 
   <div id="content"> 
      <p><b>Information retrieval</b> 
      (IR) is the science of searching for documents, for information within documents, 
      and for metadata about documents (...)
   </div>
   <div class="portal" id="p-lang"> 
      <h5>Languages</h5> (....)
   </div>
   <div id="footer">  
      <li id="footer-info-lastmod"> This page was last modified on 15 April 2011 at 23:32.</li>
      <li id="footer-info-copyright">Text is available under the 
         <a rel="license"> Creative Commons Attribution-ShareAlike License</a> 
         additional terms may apply. See Terms of Use for (....)
   </div>
</body>
\stopprog
}
\noindent A portion of the associated DOM tree is depicted in Figure~\ref{fig:DOMtree}. For the time being the reader can ignore the different colors and borders of nodes. 

\begin{figure}[h]
	\centering
		\includegraphics[width=0.8\textwidth]{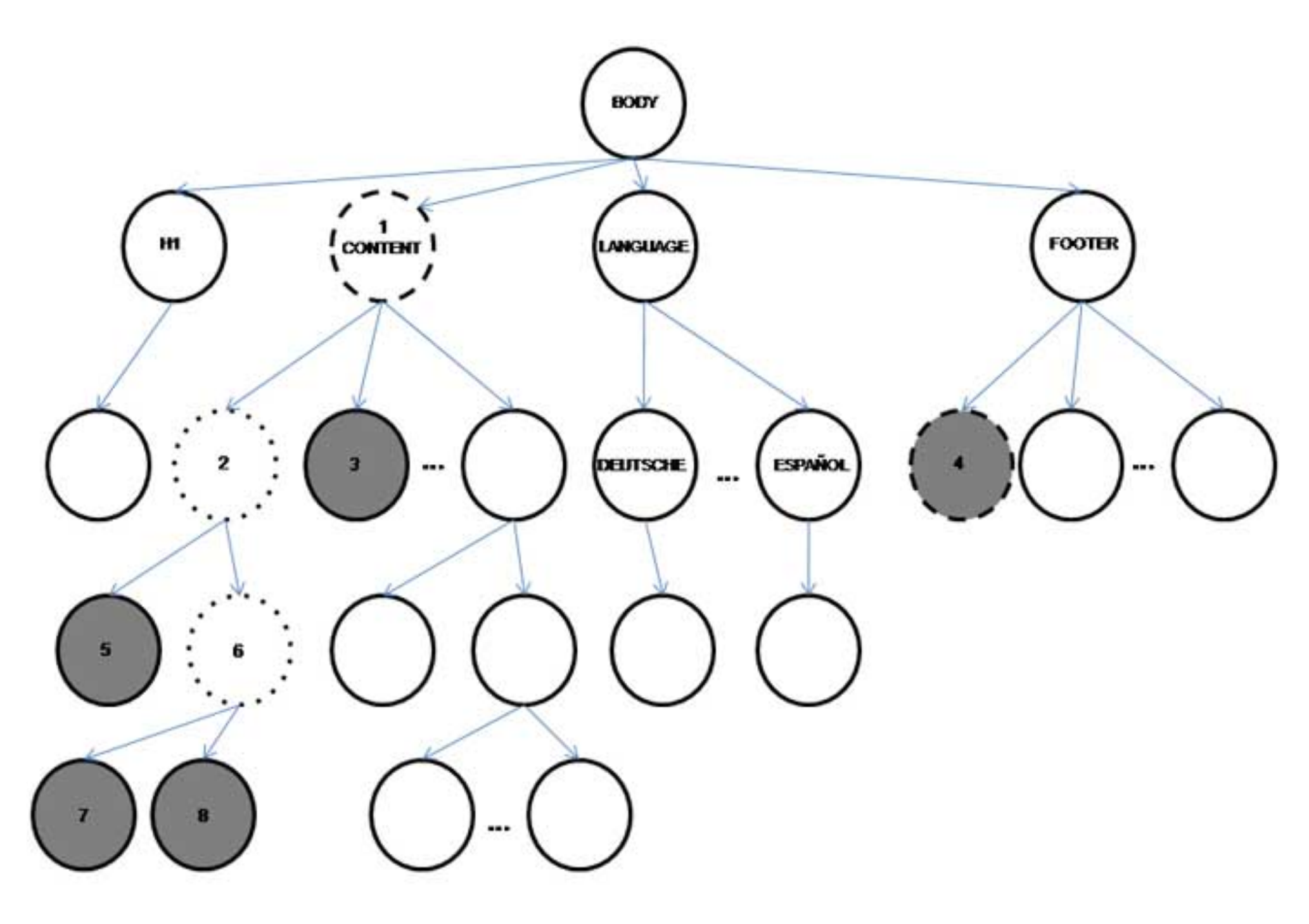}
	\caption{DOM representation of the Wikipedia's information retrieval webpage}
	\label{fig:DOMtree}
\end{figure}

\end{example}

\section{Content extraction using DOM trees}\label{sec_CE}

In this section we formalize our technique for content extraction. 
The technique is based on the notion of \emph{chars-nodes ratio} (CNR), which shows the relation between text content and tags content of each node in the DOM tree. 

\begin{definition}[chars-nodes ratio] \label{Def-ratio}
Given a DOM tree $(N,E)$, a node $n \in N$ and the set of nodes $M \subseteq N$ that form the subtree rooted at $n$ $(M=\{n' \in N~|~(n \rightarrow n')\in E^*\})$, the \emph{chars-nodes ratio} of $n$ is $chars / weight$; where $chars$ is the number of characters in the text nodes of $M$, and $weight = |M|$. 
\end{definition}

The interesting property of this definition, is that it considers nodes as blocks where the internal information is grou\-ped and indivisible using the DOM structure. Therefore, the CNR of an internal node, takes into account all the text and tags included in its descendants. Note also that the CNR of a node $n$, $\mathrm{CNR}(n)$, with a single child $n_1$ is always smaller than $\mathrm{CNR}(n_1)$ because $n$ can not contain text. However, if $n$ has several children $n_1...n_c$, then $\mathrm{CNR}(n)$ can be greater than $\mathrm{CNR}(n_1)$ depending on the amount of text in the other children. This is very useful, because it allows us to detect blocks of relevant content, even if some nodes without text belong to the block. 


Now, we are in a position to describe our method for content extraction. 
(i) We first compute the CNR for each node in the DOM tree. Then, (ii) we select those nodes
with a higher CNR and, starting from them, we traverse the DOM tree bottom-up
to find the best container nodes (e.g., tables, divs, etc.) that, roughly, contain as more relevant text as possible 
and less nodes as possible. Each of these container nodes represents an HTML block. Finally, (iii) we 
choose the block with more relevant content. All three steps can be done with a cost linear 
with the size of the DOM tree.

The first step is computed with a cost $\cO(|N|)$. With a single traversal of the tree, it ignores irrelevant code that should not be counted as text (such as Javascript), and it computes the CNRs. Even though, the computation of the CNR seems to be trivial because the DOM model's API has a method to obtain the text content of a node, this method cannot discriminate between different kinds of text contents (e.g., plain text, scripts, CSS...). Moreover, there does not exist a method to calculate the number of descendants of a given node; therefore, the computation of CNRs is done with a cumulative and recursive process that explores the DOM tree counting the text and descendants of each node. This process also allows us to detect irrelevant nodes that we call ``nonContentNode". They are, for instance, nodes without text (e.g., \emph{img}), nodes mainly used for menus (e.g., \emph{nav} and \emph{a}) and irrelevant nodes (e.g., \emph{script}, \emph{video} and \emph{svg}). This is an important advantage over other techniques that rely on the analysis of single characters or lines. These techniques cannot ignore the noisy code if they do not perform a pre-processing stage to delete these tags.

Algorithm~\ref{AlgoCNR} recursively obtains the CNR of each node starting at the root node of the DOM tree. At each node it adds three new attributes to the node with the computed weight (\emph{weight}), the number of characters it contains (\emph{textLength}), and the CNR (\emph{CNR}). The number of characters is computed ignoring special characters such as spaces or line breaks. This makes the algorithm independent of the formatting of the webpage (e.g., those webpages that organize the code using several spaces do not influence the CNRs).

\begin{algorithm}[h!]
\caption{Algorithm to compute chars-nodes ratios} \label{AlgoCNR}

\begin{algorithmic}
{\scriptsize
\smallskip
\STATE \textbf{Input:} A DOM tree $T=(N,E)$ and the root node of $T$, $root \in N$
\STATE \textbf{Output:} A DOM tree $T'=(N',E)$\\
\medskip

\STATE computeCNR(root)
\medskip

\STATE \textbf{function} ComputeCNR(node n) 
\STATE ~~\textbf{case} n.nodeType \textbf{of} \\
\STATE ~~``textNode'': 
\STATE ~~~~~~~n.addAttribute(`weight',1);
\STATE ~~~~~~~n.addAttribute(`textLength',n.innerText.length);
\STATE ~~~~~~~n.addAttribute(`CNR',n.innerText.length);
\STATE ~~~~~~~\textbf{return} n;
\STATE ~~``nonContentNode'': 
\STATE ~~~~~~~n.addAttribute(`weight',1);
\STATE ~~~~~~~n.addAttribute(`textLength',0);
\STATE ~~~~~~~n.addAttribute(`CNR',0);
\STATE ~~~~~~~\textbf{return} n;
\STATE ~~otherwise: 
\STATE ~~~~~~~descendants = 1;
\STATE ~~~~~~~charCount = 0;
\STATE ~~~~~~~\textbf{for each} child $\in$ n.childNodes \textbf{do}
\STATE ~~~~~~~~~~~newChild= ComputeCNR(child);
\STATE ~~~~~~~~~~~charCount = charCount + newChild.textLength;
\STATE ~~~~~~~~~~~descendants = descendants + newChild.weight;
\STATE ~~~~~~~n.addAttribute(`weight',descendants);
\STATE ~~~~~~~n.addAttribute(`textLength',charCount);
\STATE ~~~~~~~n.addAttribute(`CNR',charCount/descendants);
\STATE ~~~~~~~\textbf{return} n;

\smallskip

}
\end{algorithmic}
\end{algorithm}

The algorithm distinguishes between three kinds of nodes, namely \emph{textNode} which is a kind of DOM node that contains plain text and that is always a leaf, thus, it has weight 1; \emph{nonContentNode} that represents irrelevant nodes with a CNR of 0; and the rest of nodes that represent all kinds of tags. All 
methods (such as \emph{addAttribute}) and attributes (such as \emph{innerText}) used in the algorithm are standard in the DOM model and have the usual meaning.  

Once the CNRs are calculated, in the second step we select those nodes with a higher CNR. 
Then, we propagate these nodes bottom up to discover the blocks to which they belong, and the block with more text is selected. This means that if some nodes not belonging to the main block are included in the selected nodes, 
they will be discarded in the next steps.

The computation of the container blocks is performed with Algorithm~\ref{AlgoBlocks}. Roughly, this algorithm takes the DOM tree
and the set of nodes identified in the previous step, and it removes all the nodes in the set that are descendant of other nodes in the set (line (1)). Then, in lines (2) and (3), it proceeds bottom-up in the tree by discarding brother nodes and collecting their parent until a fix point is reached. This process produces a final set of nodes that represent blocks in the webpage. From all these nodes, we take the one that contains more text (in the subtree rooted at that node) as the final block.   

\begin{algorithm}[h]
\caption{Identifying main content blocks} \label{AlgoBlocks}

\begin{algorithmic}
{\scriptsize
\smallskip
\STATE \textbf{Input:} A DOM tree $T=(N,E)$ and a set of nodes $S \subset N$\\
\STATE \textbf{Output:} A set of nodes $blocks \subset N$\\
\STATE \textbf{Initialization:}
$blocks = S$\\
\bigskip

\STATE  \textbf{(1)} $blocks = blocks \backslash \{b ~|~ (b' \rightarrow b) \in E^*$ with $b,b' \in blocks\}$
\STATE  \textbf{(2)} \textbf{while} ($\exists n \in N~.~(n \rightarrow b),(n \rightarrow b') \in E$ with $b,b' \in blocks$)
\STATE  \textbf{(3)} ~~~~$blocks = (blocks \backslash \{b~|~(n \rightarrow b)\in E\}) \cup \{n\}$
\STATE
\STATE{\bf return} $blocks$
}

\smallskip

\end{algorithmic}
\end{algorithm}

\begin{example} \label{ex_algorithms}
Consider again the HTML code in Example~\ref{Ex-TagRatios} and its associated DOM tree shown in Figure~\ref{fig:DOMtree}. Algorithm~\ref{AlgoCNR} computes the CNR associated to each node of the DOM tree. All the CNRs are shown in Figure~\ref{fig:DOMCNRs}.

\begin{figure}[h]
	\centering
	       \vspace{-1.2cm}
		\includegraphics[width=0.80\textwidth]{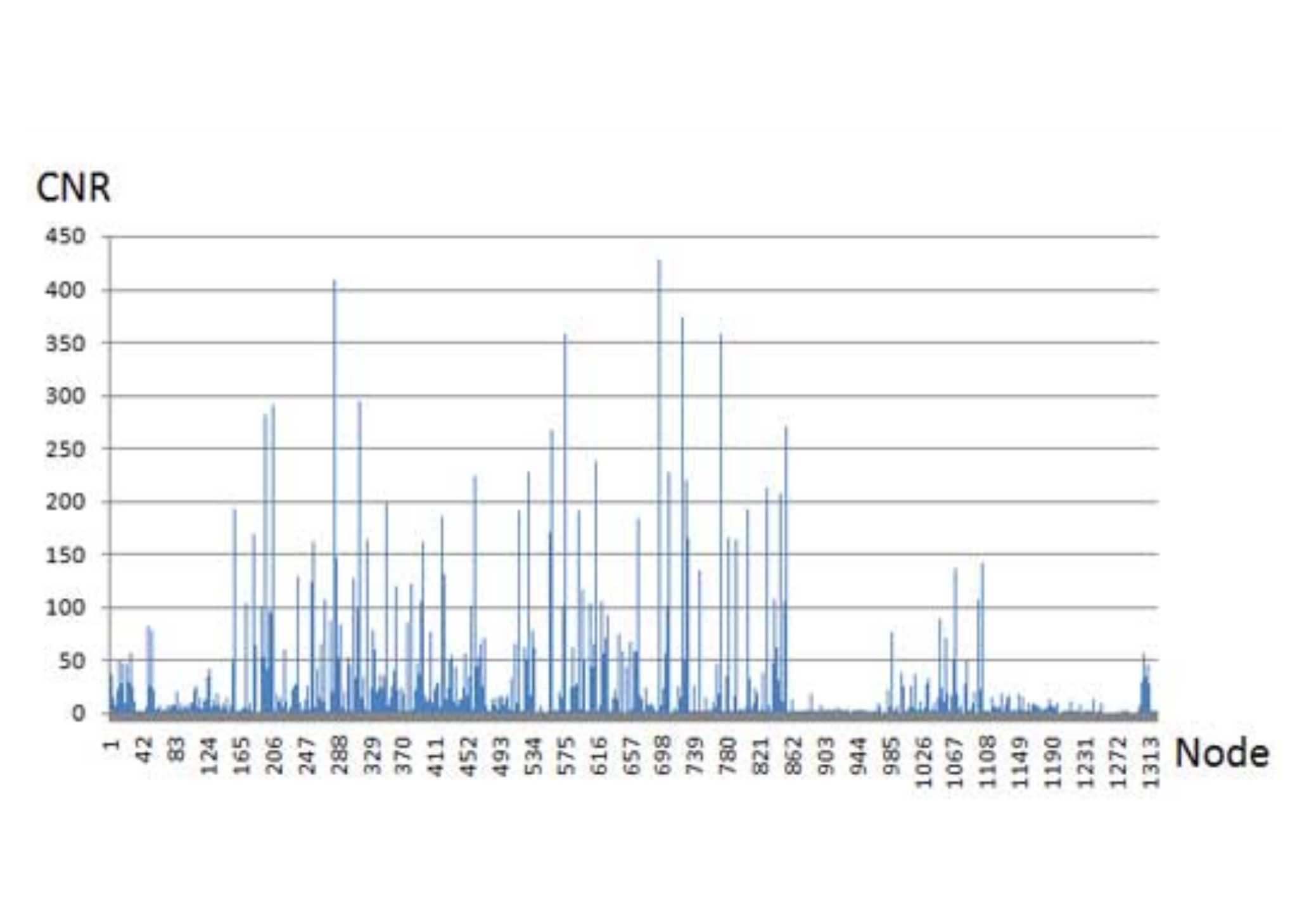}
	       \vspace{-1.2cm}
	\caption{CNRs of the DOM nodes associated with the HTML code in Example~\ref{Ex-TagRatios}}
	\label{fig:DOMCNRs}
\end{figure}

After we have computed the CNRs we take the top rated 
nodes. Let us consider that the dark nodes in Figure~\ref{fig:DOMtree} represent the top rated nodes. Then, we use Algorithm~\ref{AlgoBlocks} to identify the most relevant blocks in the webpage. Initially, all the dark nodes are in the set of blocks. Then, because nodes 7 and 8 are brothers, in the first iteration, the algorithm removes nodes 7 and 8, and it adds node 6 to the set. In the second iteration, nodes 5 and 6 are removed, and node 2 is added. Finally, in the third iteration, nodes 2 and 3 are removed, and node 1 is added. Therefore, the final set of nodes computed by Algorithm~\ref{AlgoBlocks} only contains the dashed nodes (1 and 4). The node that contains more text is selected as the block with the main content of the webpage. Observe that due to the structure of the DOM tree, the final node is often a container tag. Note also that all the nodes of this container are part of the final block, even if they do not contain text. Therefore, the final block is a block with all the information of the initial webpage that was placed together by the designer including related but non-textual elements such as images. The information of other blocks is not mixed with the information of the main block due to the structure of the DOM tree.
For instance, node 4 corresponds to the footer, and it contains a lot of text. However, although it is textually adjacent in the source code to some nodes included in the main block; it is outside the container selected as the main content block (node 1). Therefore, node 4 is finally discarded.
\end{example}

\section{Implementation} \label{sec_impl}

We have implemented the technique presented in this paper and made it publicly available, including the source code. 
It was implemented as a Firefox's plugin that can be installed in the Firefox's browser as a toolbar. Then, it can filter any loaded webpage
or produce information about the CNRs of the DOM tree.

The implementation allows the programmer to activate the transformations of the technique and to parameterize them in order to adjust the amount of blocks retrieved, and the thresholds used to detect these blocks. In order to determine the default configuration, it was initially tested with a collection of real webpages that allowed us to tune the parameters. Then, we conducted several experiments with real and online webpages to provide a measure of the average performance regarding recall, precision and the F1 measure (see, e.g., \cite{Got07} for a discussion on these metrics). 

For the experiments, we selected from the top-most 500 visited webpages (see http://www.alexa.com/topsites) a collection of domains with different layouts and page structures in order to study the performance of the technique in different contexts (e.g., company's websites, news articles, forums, etc.). Then, we randomly selected the final evaluation set.  
We determined the actual
content of each webpage by downloading it and manually 
selecting the main content text. The DOM tree of the selected text was then produced and used for comparison
evaluation later.

Table~\ref{tab:benchs} summarizes the results of the performed experiments.
The first column contains the URLs of the evaluated webpages. 
For each benchmark,
column \texttt{DOM nodes} shows the number of nodes of the whole DOM tree associated to this benchmark;
column \texttt{Main block} shows the number of nodes that were identified by the tool as the main block;
column \texttt{Recall}\ shows the number of relevant nodes retrieved divided by the total number of relevant nodes (i.e., in the main block);
column \texttt{Precision} shows the number of relevant nodes retrieved divided by the total number of retrieved nodes;
Finally, column \texttt{F1} shows the F1 metric that is computed as $(2*P*R)/(P+R)$ being $P$ the precision and $R$ the recall.

\begin{table*}[h!]
 {\footnotesize
 \centering
\begin{tabular}{|l|r|r|r|r|r|r|} \hline
 \texttt{Benchmark}~ & \texttt{DOM nodes}~ & ~\texttt{Main block}~ & ~\texttt{Recall}~ & ~\texttt{Precision}~ & {F1~~~~} \\
 \hline\hline
 \pr{www.wikipedia.org}          & 870 nodes~  & 712 nodes~  & 100 \%~ & 100 \%~  & 100 \%~ \\
 \pr{www.facebook.com}      & 744 nodes~  & 293 nodes~  & 28.6 \%~ & 100 \%~  & 44.47 \%~ \\
 \pr{www.nytimes.com}       & 742 nodes~  & 217 nodes~  & 100 \%~ & 49.7 \%~  & 66.39 \%~ \\
 \pr{www.engadget.com}          & 2897 nodes~  & 1345 nodes~  & 100 \%~ & 100 \%~  & 100 \%~ \\
 \pr{us.gizmodo.com}          & 2205 nodes~  & 1375 nodes~  & 100 \%~ & 84 \%~  & 91.3 \%~ \\
 \pr{googleblog.blogspot.com}          & 1138  nodes~  & 743 nodes~  & 100 \%~ & 100 \%~  & 100 \%~ \\
 \pr{www.bbc.co.uk}          & 401 nodes~  & 111 nodes~  & 100 \%~ & 4.98 \%~  & 9.49 \%~ \\
 \pr{www.vidaextra.com}          & 1144 nodes~  & 602 nodes~  & 100 \%~ & 100 \%~  & 100 \%~ \\
 \pr{www.gizmologia.com}          & 926 nodes~  & 415 nodes~  & 100 \%~ & 100 \%~  & 100 \%~ \\
 \pr{www.elpais.com}          & 3017 nodes~  & 120 nodes~  & 100 \%~ & 100 \%~  & 100 \%~ \\
 \pr{www.elmundo.es}          & 1722  nodes~  & 416 nodes~  & 100 \%~ & 100 \%~  & 100 \%~ \\
 \pr{www.ox.ac.uk}     & 279 nodes~  & 30 nodes~  & 100 \%~ & 28 \%~  & 43.75 \%~ \\
 \pr{www.thefreedictionary.com}  & 1170  nodes~  & 509 nodes~  & 100 \%~ & 100 \%~  & 100 \%~ \\
 \pr{www.nlm.nih.gov}       & 320 nodes~  & 156 nodes~  & 100 \%~ & 56.52 \%~  & 71.56 \%~ \\
 \pr{www.scielosp.org}      & 563  nodes~  & 458 nodes~  & 100 \%~ & 100 \%~  & 100 \%~ \\
 \pr{www.wordreference.com}   & 269  nodes~  & 95 nodes~  & 100 \%~ & 57.23 \%~  & 72.79 \%~ \\
 \pr{en.citizendium.org}      & 1645  nodes~  & 1478 nodes~  & 100 \%~ & 100 \%~  & 100 \%~ \\
 \pr{knol.google.com}          & 601 nodes~  & 219 nodes~  & 100 \%~ & 100 \%~  & 100 \%~ \\
 \pr{www.healthopedia.com}     & 557 nodes~  & 21 nodes~  & 100 \%~ & 21 \%~  & 34.7 \%~ \\
 \pr{www.filmaffinity.com}          & 1198  nodes~  & 153 nodes~  & 100 \%~ & 100 \%~  & 100 \%~ \\
 \pr{www.umm.edu}          & 290 nodes~  & 30 nodes~  & 100 \%~ & 22.22 \%~  & 36.42 \%~ \\
 \pr{www.microsiervos.com}          & 604    nodes~  & 382 nodes~  & 100 \%~ & 68.83 \%~  & 81.54 \%~ \\
 \pr{abcnews.go.com}          & 907   nodes~  & 102 nodes~  & 100 \%~ & 44.16 \%~  & 61.27 \%~ \\
 \pr{www.latimes.com}          & 1056   nodes~  & 22 nodes~  & 100 \%~ & 100 \%~  & 100 \%~ \\
 \pr{www.philly.com}          & 378   nodes~  & 30 nodes~  & 100 \%~ & 100 \%~  & 100 \%~ \\
 \pr{www.blogdecine.com}          & 1567    nodes~  & 24 nodes~  & 100 \%~ & 8.33 \%~  & 15.38 \%~ \\
 \pr{www.cnn.com}          & 597   nodes~  & 248 nodes~  & 100 \%~ & 67.21 \%~  & 80.39 \%~ \\
 \pr{www.lashorasperdidas.com}          & 87   nodes~  & 30 nodes~  & 100 \%~ & 100 \%~  & 100 \%~ \\
 \pr{www.cbc.ca}          & 847    nodes~  & 138 nodes~  & 100 \%~ & 100 \%~  & 100 \%~ \\
 \pr{www.appleweblog.com}          & 1013   nodes~  & 475 nodes~  & 5.9 \%~ & 100 \%~  & 11.15 \%~ \\
 \pr{www.applesfera.com}          & 1215    nodes~  & 721 nodes~  & 7.49 \%~ & 100 \%~  & 13.94 \%~ \\
 \pr{www.guardian.co.uk}          & 1111   nodes~  & 59 nodes~  & 100 \%~ & 100 \%~  & 100 \%~ \\
 \pr{www.news.cnet.com}          & 2023    nodes~  & 169 nodes~  & 100 \%~ & 71.01 \%~  & 83.05 \%~ \\
 \pr{www.venturebeat.com}          & 263   nodes~  & 107 nodes~  & 100 \%~ & 100 \%~  & 100 \%~ \\
 \pr{www.computerworld.com}          & 558   nodes~  & 62 nodes~  & 100 \%~ & 100 \%~  & 100 \%~ \\
 \pr{www.usatoday.com}          & 1118    nodes~  & 523 nodes~  & 100 \%~ &  100 \%~  & 100 \%~ \\
 \pr{www.cbssports.com}          & 1450   nodes~  & 232 nodes~  & 100 \%~ & 67.05 \%~  & 80.28 \%~ \\
 \pr{www.nationalfootballpost.com}          & 565    nodes~  & 23 nodes~  & 100 \%~ & 9.62 \%~  & 17.55 \%~ \\
 \pr{ncaabasketball.fanhouse.com}          & 885    nodes~  & 78 nodes~  & 100 \%~ & 40.20 \%~  & 57.35 \%~ \\
 \pr{www.sportingnews.com}          & 1394    nodes~  & 79 nodes~  & 100 \%~ & 72.48 \%~  & 84.05 \%~ \\
 \pr{www.hoopsworld.com}          & 629   nodes~  & 112 nodes~  & 100 \%~ & 100 \%~  & 100 \%~ \\
 \pr{profootballtalk.nbcsports.com}          & 394   nodes~  & 28 nodes~  & 100 \%~ & 45.17 \%~  & 62.23 \%~ \\
 \pr{www.thehollywoodgossip.com}          & 362   nodes~  & 44 nodes~  & 100 \%~ & 100 \%~  & 100 \%~ \\
 \pr{www.rollingstone.com}          & 993    nodes~  & 29 nodes~  & 100 \%~ & 20.42 \%~  & 33.92 \%~ \\
 \pr{popwatch.ew.com}          & 919   nodes~  & 93 nodes~  & 100 \%~ & 100 \%~  & 100 \%~ \\
 \pr{www.people.com}          & 923   nodes~  & 56 nodes~  & 100 \%~ & 32 \%~  & 48.49 \%~ \\
 \pr{www.cinemablend.com}          & 495   nodes~  & 59 nodes~  & 100 \%~ & 37.34 \%~  & 54.38 \%~ \\ \hline
\end{tabular}
}
\caption{Benchmark results}
\label{tab:benchs}
\end{table*}

Experiments reveal that in many cases, the retrieved block is exactly the relevant block (F1=100\%), and in general, the recall is 100\%. This means that the retrieved block often contains all the relevant information. 
The average recall is 94.39 and the average precision is 74.08. These are really good measures. For instance, with the same webpages, the best previous technique (using tag ratios \cite{Wen10}) produces an average recall of 92.72 and an average precision of 71.93.

Observe one important property of the experiments: in all cases, either the recall, the precision, or both, are 100\%. This phenomenon did not happen by a chance, it is a direct consequence of the way in which the technique selects blocks. Let us consider a DOM tree where node $n$ is the actual relevant block. Our technique explores the DOM tree bottom-up to find this node, and only four cases are possible: (1) If we detect node $n$ as the main block, then both recall and precision are 100\%. (2) If we choose a node that is a descendant of $n$, then precision is 100\%. (3) If we choose a node that is an ancestor of $n$, then recall is 100\%. Finally, (4) if we select a node that is not an ancestor neither a descendant of $n$ then both recall and precision would be 0\%. This case is very rare because this would mean that there exists a non-relevant block that contains more text than the relevant block. This never happened in all the experiments we did. 

We could take advantage of this interesting characteristic of our technique. We could parameterize the technique to ensure that we have a 100\% recall, or to ensure that we have a 100\% precision depending on the applications where it is used. This can be easily done by making Algorithm~\ref{AlgoBlocks} to be more restrictive (i.e., selecting blocks closer to the leaves, thus, ensuring 100\% precision), or more relaxed (i.e., selecting blocks closer to the root, thus, ensuring 100\% recall). 

All the information related to the experiments, the source code of
the benchmarks, the source code of the tool
and other material can be found at \url{http://users.dsic.upv.es/~jsilva/CNR}


%



\section{Cases of Study}
\label{sec_CS}

This section provides three real examples of content extraction using our technique. The scenarios have been selected to show what is the information that the user gets when the technique works well, when it includes more information than needed, and when it loses some information.

\bigskip
\noindent{\bf Content extraction from Wikipedia}\\

\begin{figure*}[h!]
	\centering
		\includegraphics[width=0.48\textwidth]{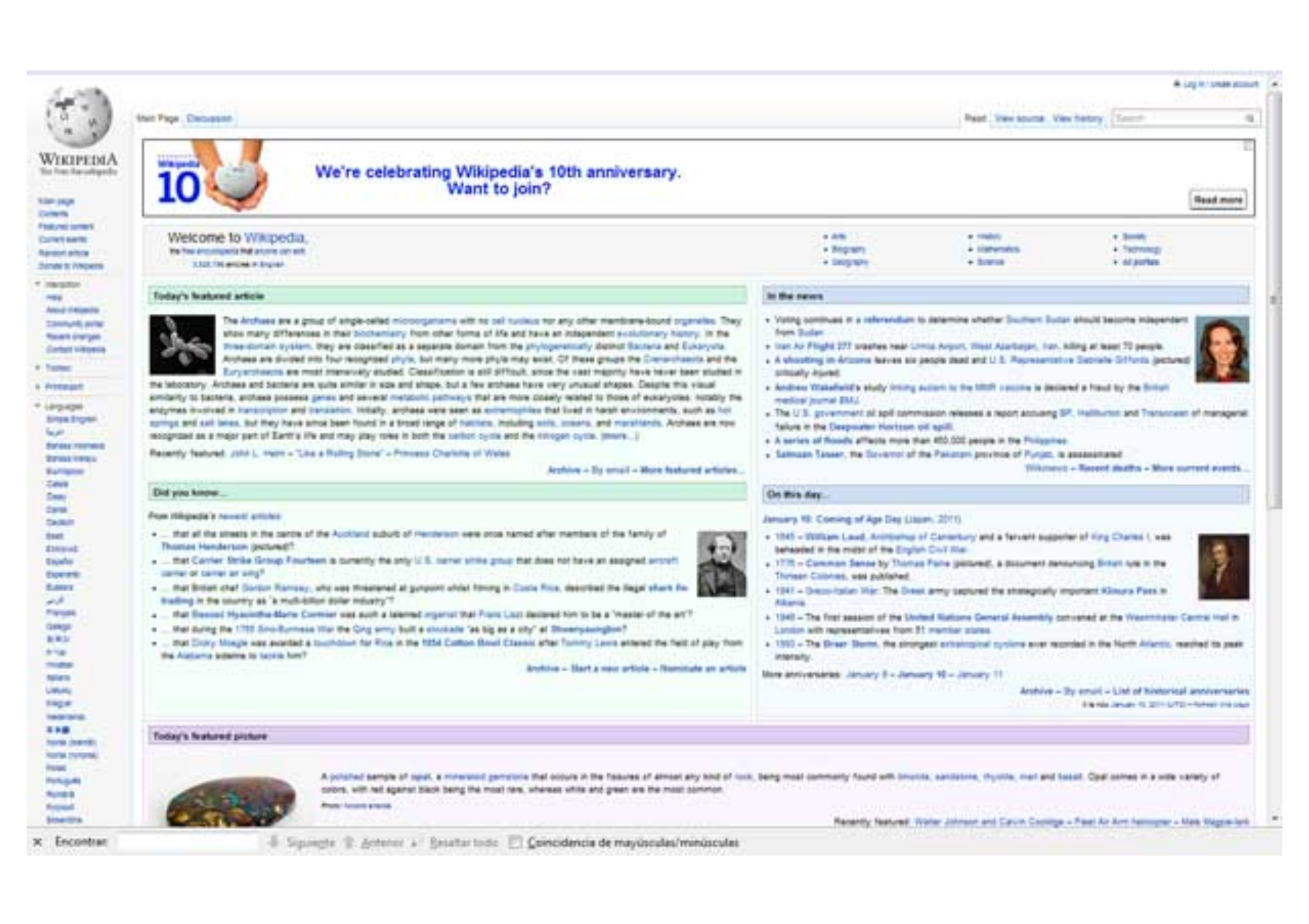}~~~
		\includegraphics[width=0.48\textwidth]{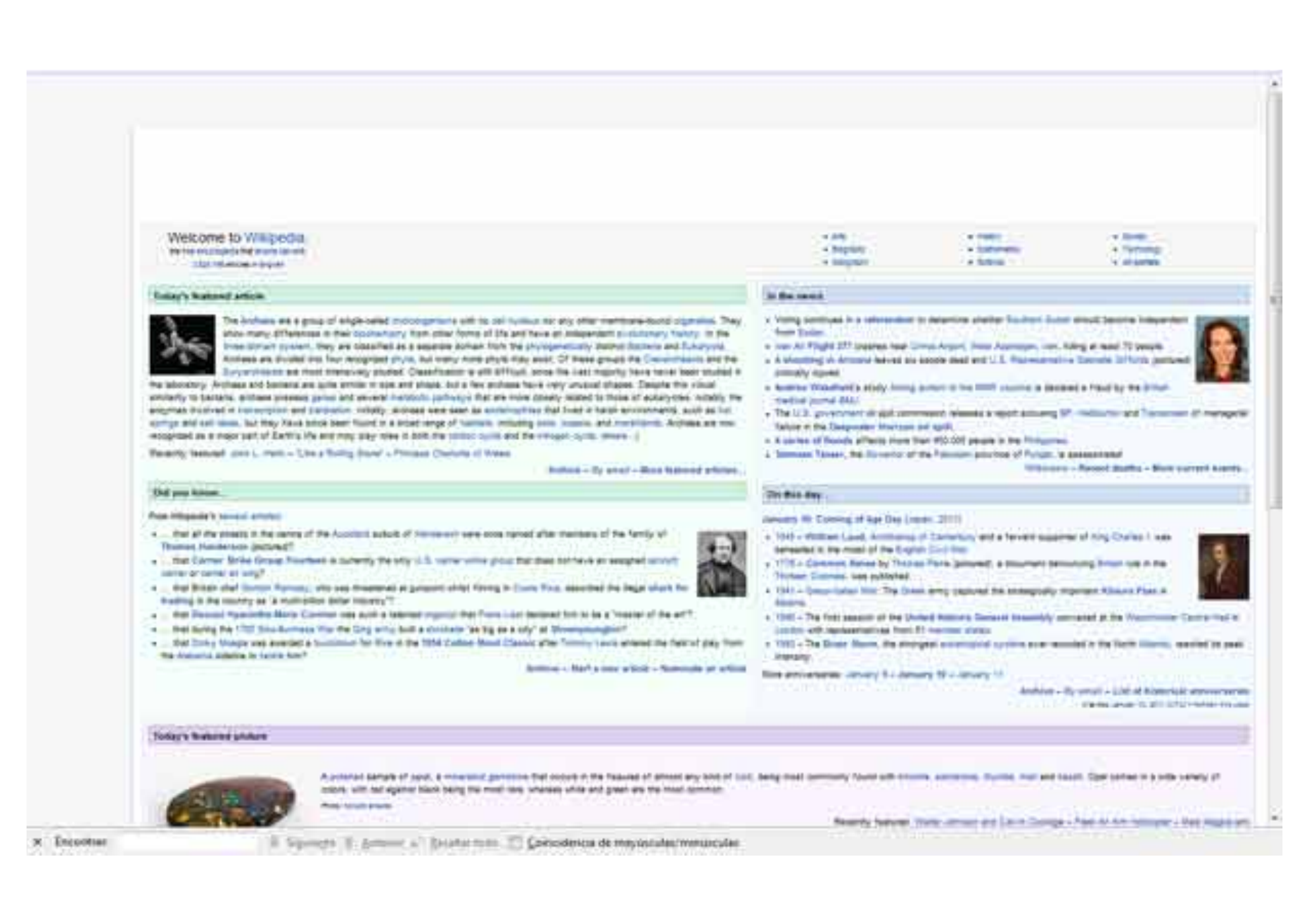}
	\caption{Main webpage of Wikipedia (left) and its filtered version (right)}
	\label{fig:wikipedia}
\end{figure*}

The first scenario is the most common situation. It happens when both the precision and the recall are 100\%, thus the content extracted is exactly the main content. 
This situation happens around 50\% of the times. As an example, observe in Figure~\ref{fig:wikipedia} that the main webpage of Wikipedia has been filtered and both the horizontal and vertical menus have been removed. All the advertisements have been also removed. 

\bigskip
\noindent{\bf Content extraction from Filmaffinity}\\

The recall of the technique is 100\% in 94\% of the cases. This means that the main content is normally recovered. However, sometimes the technique recovers some extra content. This extra content is usually around the main content and sometimes is a layer that the designer of the webpage inserted in the middle of the main content. It often shows a picture or video related to the main content, but sometimes it is just an advertisement. This is clearly a bad design policy used by the programmer that avoids to correctly reuse the code. However, we believe that it is done on purpose to ensure that those webpages that reuse the main content will force the user to see the advertisements.

\begin{figure*}[h!]
	\centering
	       \vspace{-1.5cm}
		\includegraphics[width=0.48\textwidth]{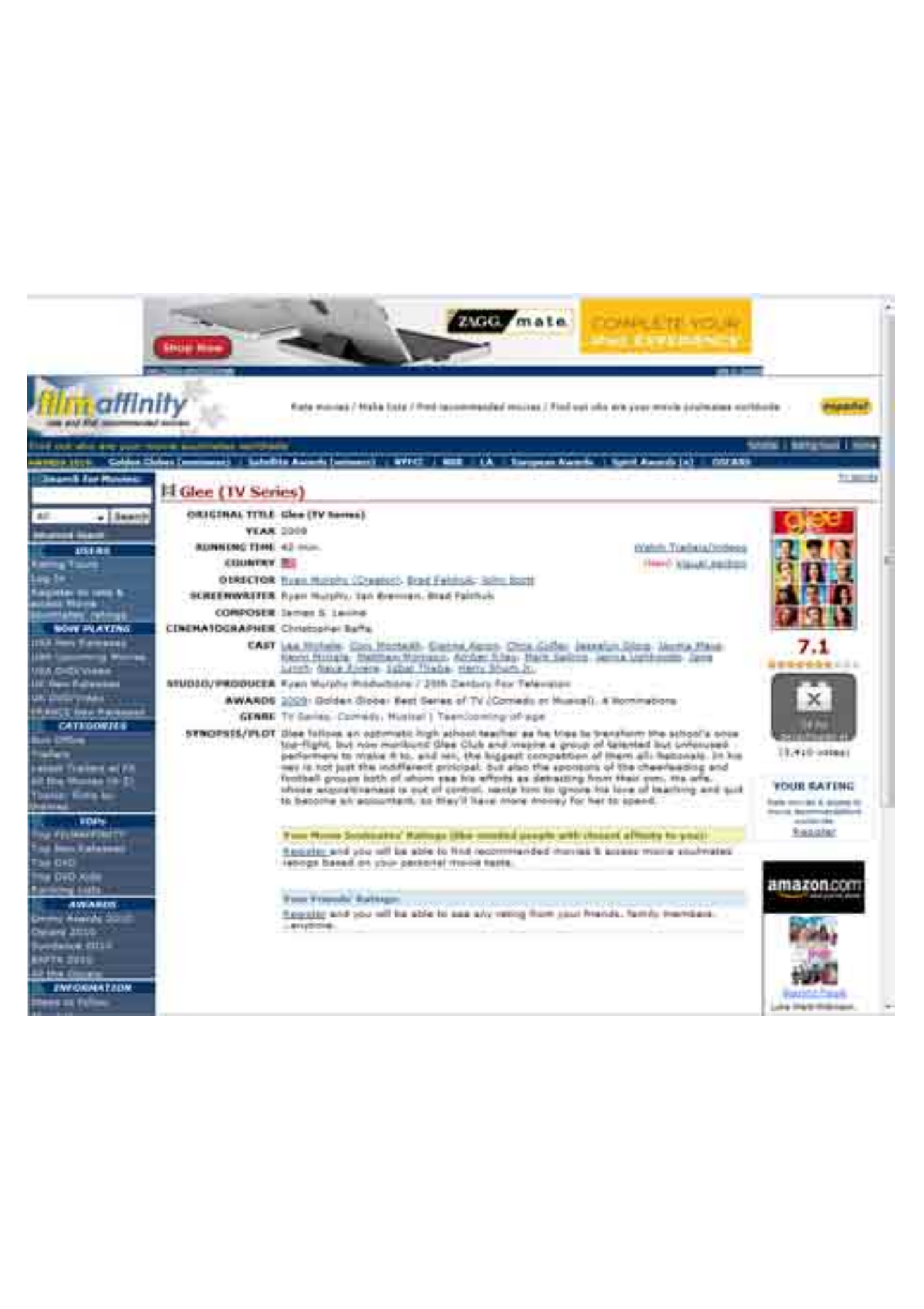}~~~
		\includegraphics[width=0.48\textwidth]{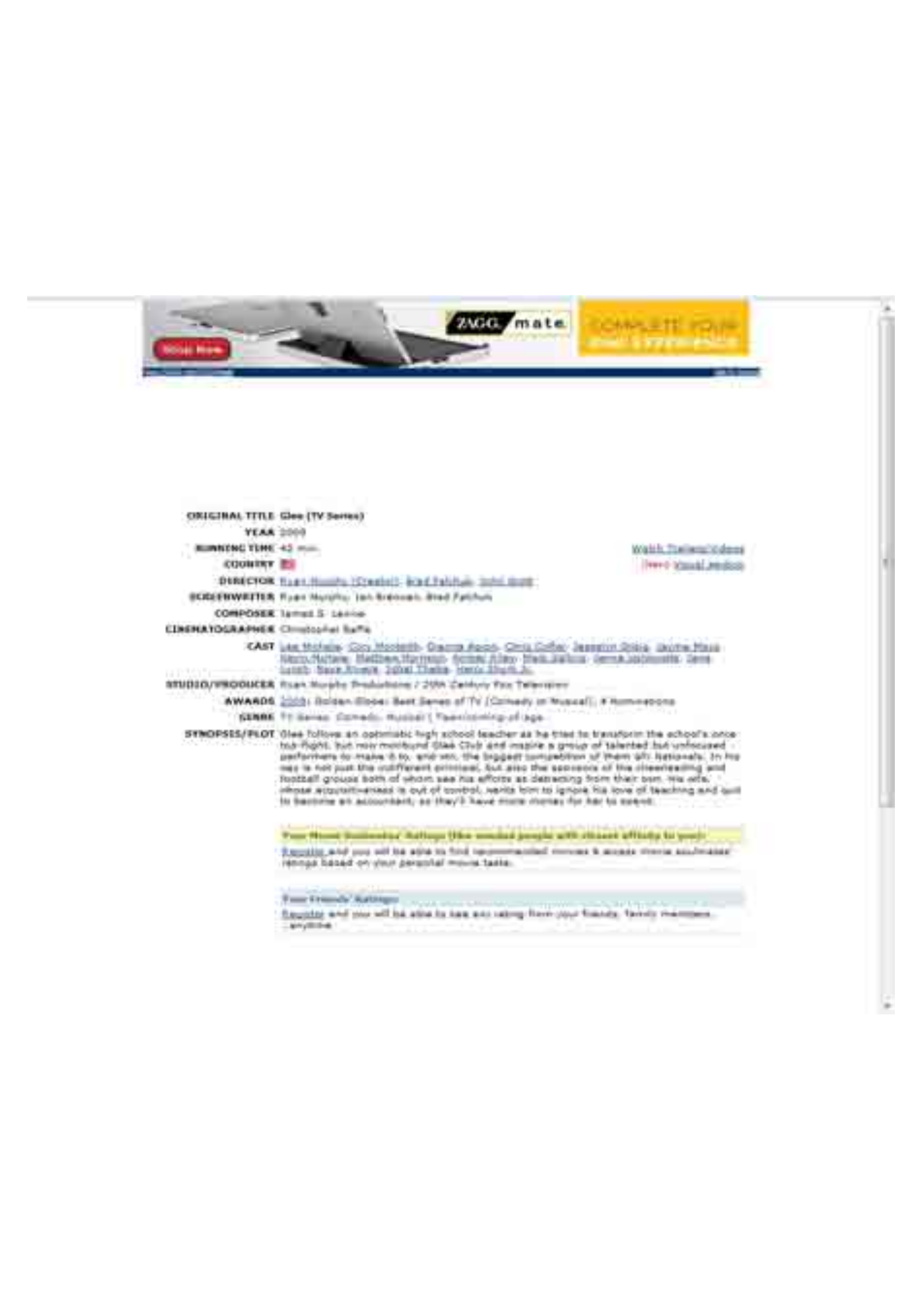}
	       \vspace{-1.9cm}		
	\caption{Filmaffinity's webpage (left) and its filtered version (right)}
	\label{fig:filmaffinity}
\end{figure*}

We can observe this phenomenon in Figure~\ref{fig:filmaffinity}. Observe how this webpage taken from Filmaffinity has been almost perfectly filtered. Everything except the main content has been removed with one exception: the top advertisement panel. In those experiments where the tool did include more information than needed, it was usually caused by this situation. It usually included the main content and some additional container (usually a \emph{div} label) with extra information. 

\bigskip
\noindent{\bf Content extraction from BBC}\\

\begin{figure*}[h!]
	\centering
	       \vspace{-1.3cm}	
		\includegraphics[width=0.48\textwidth]{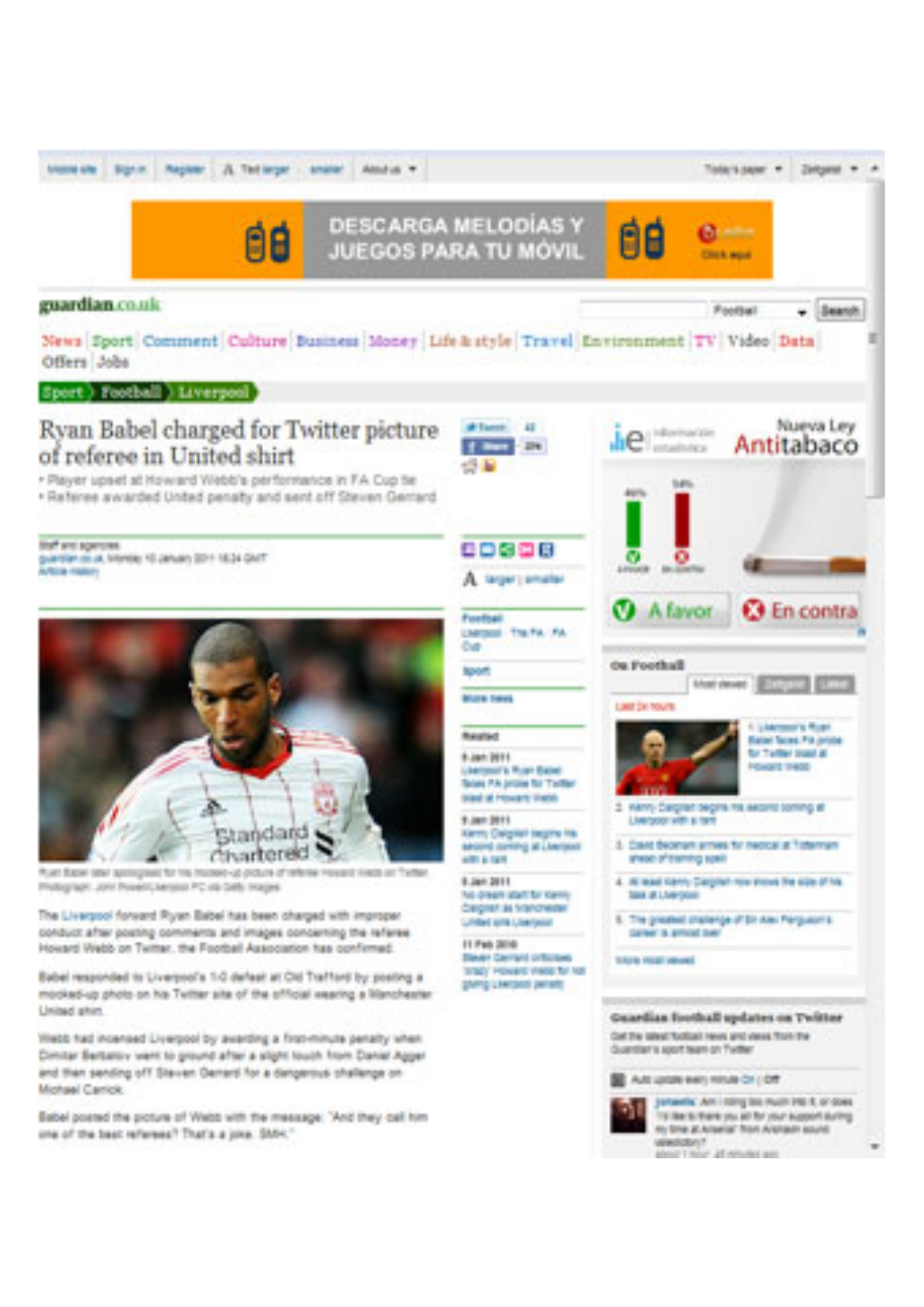}~~~
		\includegraphics[width=0.48\textwidth]{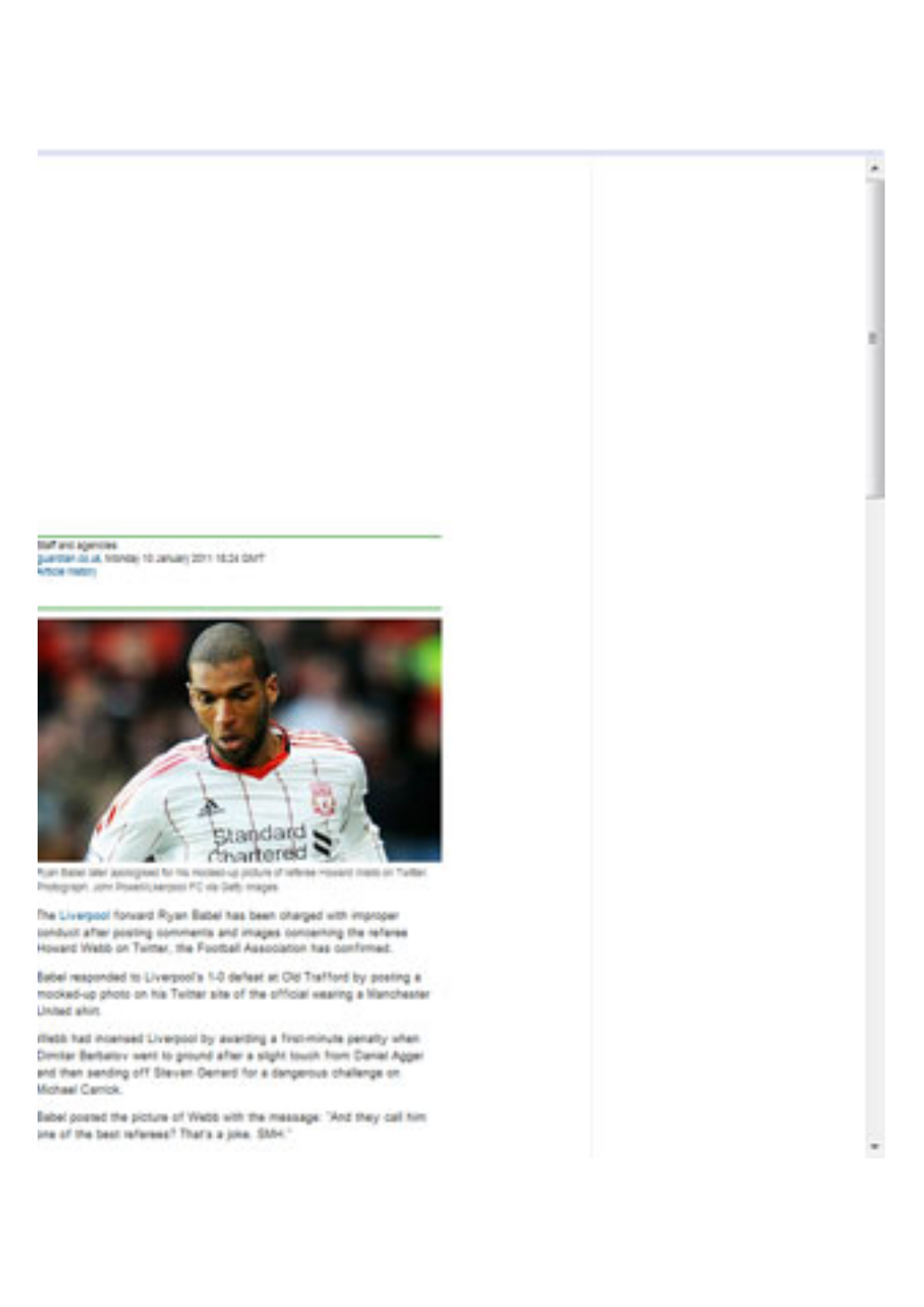}
	       \vspace{-1.1cm}		
	\caption{News article at BBC's webpage (left) and its filtered version (right)}
	\label{fig:bbc}
\end{figure*}

It is difficult that the technique loses some information from the main content. This only happens 6\% of the times. 
An example of this scenario is shown in Figure~\ref{fig:bbc}. Observe that the title of the article is missing in the content extracted.
In order to assist the user in these rare cases, we added two buttons to the tool that allows her to increase or decrease the amount of information shown. In this case, by clicking on the `increase information' button, the whole main content would be perfectly displayed.

The implementation of these buttons is trivial, but they are really useful to provide the user with control over the retrieved information.
As explained before, the main content extracted can be represented with a single node of the DOM tree (i.e., this node and all the descendants are the main content). Increasing the amount of information shown is done by selecting the parent of this node as the new main content. Observe that the new node is necessarily an HTML container. Analogously, decreasing the information shown is done by selecting the child node with a higher CNR.
 
\section{Conclusions}
\label{sec_concl}

Content extraction is useful not only for the final user, but also for many systems and tools such as indexers as a preliminary stage. It extracts the relevant part of a webpage allowing us to ignore the rest of content that can become useless, irrelevant, or even worst, noisy. 
In this work, we have presented a new technique for content extraction 
uses the DOM structure of the webpage to identify the blocks that groups those nodes with a higher proportion of text.

The DOM structure allows us to improve the detection of blocks, but it also allows us to discard parts of the webpage that have a big amount of textual information but belong to other HTML containers. Our implementation and experiments have shown the usefulness of the technique.

The technique could be used not only for content extraction, but also for blocks detection. It could detect all blocks in a webpage by applying the presented algorithms iteratively to detect one block after the other. In this way, we could detect the most relevant block; then, remove from the DOM tree all its nodes, and detect the next relevant block in the remaining DOM tree. This process would identify all blocks in relevant order. 
Another interesting open line of research is using the technique to detect the menus of a webpage. A preliminary study showed that instead of using a ratio characters/nodes, we could use a ratio hyperlinks/nodes to discover big concentrations of links in the DOM tree. If we collect those concentrations of links where the links contain less characters, we will find the menus of the webpage.


\begin{thebibliography}{}
\providecommand{\urlalt}[2]{\href{#1}{#2}}
\providecommand{\doi}[1]{doi:\urlalt{http://dx.doi.org/#1}{#1}}

\bibitem{Bal06}
S.~Baluja.
\newblock Browsing on small screens: Recasting web-page segmentation into an
  efficient machine learning framework.
\newblock In {\em Proceedings of the 15th International Conference on World
  Wide Web (WWW'06)} pages 33--42, New York, NY, USA, 2006. ACM.
  \doi{10.1145/1135777.1135788}

\bibitem{Coh02}
W.~W. Cohen, M.~Hurst, and L.~S. Jensen.
\newblock A flexible learning system for wrapping tables and lists in html
  documents.
\newblock In {\em In Proceedings of the international World Wide Web conference
  (WWW'02)}, pages 232--241, 2002.
  \doi{10.1145/511475.511477}

\bibitem{Dal11}
B.~Dalvi, W.~W. Cohen, and J.~Callan.
\newblock Websets: Extracting sets of entities from the web using unsupervised
  information extraction.
\newblock Technical report, Carnegie Mellon School of computer Science, 2011.
  \doi{10.1145/2124295.2124327}

\bibitem{Gib07}
J.~Gibson, B.~Wellner, and S.~Lubar.
\newblock Adaptive web-page content identification.
\newblock In {\em Proceedings of the 9th annual ACM international workshop on
  Web information and data management}, WIDM '07, pages 105--112, New York, NY,
  USA, 2007. ACM.
  \doi{10.1145/1316902.1316920}

\bibitem{Koh09}
C.~Kohlsch\"{u}tter.
\newblock A densitometric analysis of web template content.
\newblock In {\em Proceedings of the 18th international World
  Wide Web conference (WWW'09)}, pages 1165--1166, New York, NY, USA, 2009. ACM.
  \doi{10.1145/1526709.1526909}

\bibitem{Koh10}
C.~Kohlsch\"{u}tter, P.~Fankhauser, and W.~Nejdl.
\newblock Boilerplate detection using shallow text features.
\newblock In {\em Proceedings of the third ACM international conference on Web
  search and data mining}, WSDM '10, pages 441--450, New York, NY, USA, 2010.
  ACM.
  \doi{10.1145/1718487.1718542}

\bibitem{Koh08}
C.~Kohlsch\"{u}tter and W.~Nejdl.
\newblock A densitometric approach to web page segmentation.
\newblock In {\em Proceeding of the 17th ACM conference on Information and
  knowledge management}, CIKM '08, pages 1173--1182, New York, NY, USA, 2008.
  ACM.
  \doi{10.1145/1458082.1458237}

\bibitem{Kus97}
N.~Kushmerick, D.~S. Weld, and R.~Doorenbos.
\newblock Wrapper induction for information extraction.
\newblock In {\em Proceedings of the Fifteenth International Joint Conference
  on Artificial Intelligence (IJCAI'97)}, 1997.

	\bibitem{Gib05} D. Gibson, K. Punera, and A. Tomkins. The volume and evolution of web page templates. In Proceedings of the 14th International Conference on World Wide Web (WWW'05). pp. 830-839, Chiba, Japan, 
	\doi{10.1145/1062745.1062763}
	\bibitem{Li03} X. Li and B. Liu. Learning to classify text using positive and unlabeled data. In Proceedings of the International Joint Conference on Artificial Intelligence (IJCAI'03), Acapulco, Mexico, 2003.
	\bibitem{Ari09} J. Arias, K. Deschacht and M.F. Moens. Language independent content extraction from web pages. In Proceedings of the 9th Dutch-Belgian Information Retrieval Workshop (DIR'09), pp. 50-55, The Netherlands, 2009.
	\doi{ 10.1145/1299015.1299021}
	\bibitem{KHG05} B. Kr\"upl, M. Herzog, and W. Gatterbauer. Using visual cues for extraction of tabular data from arbitrary HTML documents. In Proceedings of the 14th International Conference on World Wide Web (WWW'05), Chiba, Japan, 2005.
	\doi{10.1145/1062745.1062838}
	\bibitem{Got08} T. Gottron. Content code blurring: A new approach to content extraction. In Proceedings of the 5th International Workshop on Text-Based Information Retrieval (TIR'08), pp. 29-33, Turin, Italy, 2008.
	\doi{10.1109/DEXA.2008.43}
	\bibitem{Wen10} T. Weninger, W.H. Hsu and J. Han. CETR - Content Extraction via Tag Ratios. In Proceedings of the 19th International Conference on World Wide Web (WWW'10), pp. 971-980, North Carolina, USA, 2010.
	\doi{ 10.1145/1772690.1772789}
	\bibitem{Gup03} S. Gupta, G. Kaiser, D. Neistadt and P. Grimm. DOM-based content extraction of HTML documents. In Proceedings of the 12th International Conference on World Wide Web (WWW'03), pp. 207-214, North  Budapest, Hungary, 2003.
	\doi{10.1145/775181.775182}
	\bibitem{Fin01} F. Finn, N. Kushmerick and B. Smyth. Fact or fiction: Content classification for digital libraries. DELOS-NSF Workshop on Personalisation and Recommender Systems in Digital Libraries, Dublin, 2001.
        \bibitem{DOM} W3C Consortium, Document Object Model (DOM). www.w3.org/DOM
	\bibitem{Got07} T. Gottron. Evaluating content extraction on HTML documents. In Proceedings of the 2nd International Conference on Internet Technologies and Applications (ITA'07), pp. 123-132, Wrexham, North Wales, 2007.

%
\end{thebibliography}
\nocite{*}
\bibliographystyle{eptcs}

\end{document}